\documentstyle[12pt]{article}

\topmargin = - 2.5 cm
\begin{document}

\renewcommand
\baselinestretch{1.5}
\baselineskip=2em

\title{\large \bf GRAVITATIONAL FORCES WITH STRONGLY LOCALIZED
RETARDATION}
\author{ \normalsize Thomas Chen \thanks{e-mail: chen@ifm.mavt.ethz.ch}
\and  \normalsize William D. Walker \thanks{e-mail:
walker@ifm.mavt.ethz.ch}\and   \normalsize J\"{u}rg Dual \thanks{e-mail:
dual@ifm.mavt.ethz.ch}}
\date{\small {\em Institute of Mechanics, Swiss Federal Institute of
Technology}
\\
\small {\em 8092 Z\"{u}rich, Switzerland} }
\maketitle

\vspace{1cm}

\begin{abstract}  We solve the linearized Einstein equations for a
specific oscillating mass  distribution and discuss the usual
counterarguments against the existence of observable gravitational
retardations  in the "near zone", where $\frac{d}{r}\ll 1$ ($d=$
oscillation amplitude of the source, $r=$ distance from the source).
We  show that they do not apply in the region
$\frac{d}{r}\approx 1$, and prove that gravitational forces are retarded
in the immediate vicinity of the source.  An experiment to measure this
retardation is proposed, which may provide the first direct experimental
observation of propagating gravitational fields.
\end{abstract}

\vspace{2.5cm} PACS numbers: 04.30.-w, 04.80.Nn, 04.80.Cc, 04.80.-y

\section{\large \bf INTRODUCTION AND SURVEY OF RESULTS}

Between the 1940s and the early 1960s, several researchers proposed the
possibility of measuring the retardation of gravitational interaction
using laboratory  experiments
$[1,2,3,4]$. Because of the poor technology at that time and
contradictory theoretical predictions about the retardation,  interest in
the subject declined. No experiments were performed during this  period,
and after 1962, the topic was never discussed again in the literature.
At the 1996  Virgo conference, a new experimental set-up was described
for the purpose of  testing the retardation of gravitational interaction
in the immediate vicinity of  a vibrating mass distribution $[5]$.
Discussions following the talk indicated that there is still ambiguity
about the existence of such a retardation. In this paper, we will discuss
the common arguments which lead to the conclusion that gravitational
forces generated by  oscillating mass distributions are instantaneous in
the near-field. All of these arguments have the weakness that they are
not valid at extremely small distances from the oscillating source. We
will analyze the gravitational forces in the immediate vicinity of a
specific oscillating mass distribution, for which the  dominant  part of
the corresponding linearized Einstein equations can be solved.  It shows
that there is a possibly detectable retardation of the gravitational
forces in the immediate vicinity of the source, which converges to zero
with increasing  distance. The dominant length scale for the decay of the
retardation is given by  the {\em oscillation amplitude} of the mass
distribution and can be very small. Experimental verification of this
local retardation effect would provide the first direct observation of
propagating gravitational fields. We propose a laboratory experiment
which requires a highly accurate phase measurement to test our  result.
Given the state of modern  technology, this measurement may now be
possible.

\section{\large \bf DEFINITION, DISCUSSION AND ANALYSIS OF THE  \\
 MATHEMATICAL MODEL}

We consider an idealized experiment in a laboratory environment,
consisting of a mass-balanced, mechanical two-body oscillator, which
generates disturbances in the space-time metric. The oscillator consists
of a point mass $m$ which  is connected by way of a spring with
negligible mass to a second point mass $M
\gg m$. The resulting mass-spring system oscillates with eigenfrequency
$\omega$ about its center of gravity, and does not interact dynamically
with additional, external masses. The distance between the masses
$m$ and $M$ is assumed to be much larger than the oscillation amplitude
$d$ of the smaller mass $m$. The oscillation amplitude $D=\frac{m}{M}d$
of $M$, which is calculated with classical mechanics, is much smaller
than $d$ and leads to negligible time-dependent gravitational effects in
the immediate vicinity of
$m$, as will be shown in section II.1. We will mainly consider the
time-dependent gravitational forces generated by
$m$, and will thus refer to $m$ as the "source", or the "source mass".
Another mechanical oscillator,  tuned to resonance to detect these
gravitational forces, is placed on the oscillator axis at a distance of
order
$O(d)$ from $m$. The purpose of this paper is to show that the
gravitational forces generated in this extreme near-field  laboratory
experiment are retarded, which may be verifiable with modern technology.
Due to many objections against experiments of this kind in the past, we
will first discuss the common counterarguments.

The occurrence of a measurable retardation in the proposed system may
seem  problematic  for several reasons. One may suspect that the theorem
about the  non-existence of  gravitational dipole radiation is violated
by the mass-balanced, mechanical two-body oscillator described above. It
states that isolated mechanical systems  cannot generate any dipolar
gravitational radiation, due to momentum and angular  momentum
conservation within the system
$[6]$. However, this argument does not contradict the existence of
retarded gravitational forces which are strongly  localized around the
source. We will refer to the {\em immediate vicinity} of $m$ as a region
around $m$ with a diameter of order $O(d)$. In section II.1, we will
prove that  the {\em time-dependent} gravitational forces due to the
larger mass $M$ can be neglected in the immediate vicinity of the source
mass $m$, although the static field due to $M$ may dominate over the
static field generated by $m$. The mass $m$, which is by itself not an
isolated system,  appears locally to be an isolated oscillator.  Because
the momentum of $m$ is by itself not conserved, $m$ is allowed to radiate
gravitationally in its immediate vicinity without contradicting the
overall conservation of momentum. At larger distances from the
oscillator, the time-dependent gravitational forces due to $M$ and $m$
have equal strength, and we will show  in section II.2.1 that their
radiative contributions cancel. Hence, there is no contradiction to
overall momentum  conservation.

Near-field set-ups similar to the proposed one are usually analyzed with
the "near  zone" limit used in radiation theory,  typically defined as a
distance range comparable to the wavelength of the radiated signals,
which is usually very large. At these large distances  from the source,
the vibrating mass distribution can be very accurately modelled as a
series of multipole coefficients, which neglect all retardation effects
within the source. But in the immediate vicinity of the source, i.e. at
distances comparable to its vibration amplitude, it is not possible to
make this simplification. Instead, one must take local retardation
effects into account, even if they disappear at large distances.

Another common objection against  the existence of observable
retardations in the proposed set-up arises from an electrodynamic
analogy. Retarded electromagnetic  potentials may result in almost
instantaneous electromagnetic fields, due to certain  cancellation
effects $[7]$. This phenomenon occurs in the "near zone" introduced
above, and is consistent with our results. We will show in section II.2.2
that the retardation of the gravitational forces decreases {\em
continuously} to zero with increasing distance from the source, and that
in the immediate vicinity of $m$, the retardation has possibly measurable
values.
\\

We will first show that our proposed system can be analyzed with the
linearized Einstein equations.  Let $c \approx 3 \cdot
10^{8}[\frac{m}{sec}]$ denote the  vacuum speed of light. We choose a
local Minkowskian coordinate frame $\lbrace ct,x,y,z \rbrace$, in which
$m$ oscillates about the origin with amplitude $d$, angular frequency
$\omega$,  and parallel to the $z$ axis
\begin{equation}
\vec{z}(t) = \; \hat{z} \; d \: \sin \omega t
\end{equation}
$\hat{z}$ denotes the unit vector in the $z$ direction, and $\vec{z}(t)$
is the location of $m$ at time $t$. We assume mass $m \approx 
O(10^{-1}[kg])$,  oscillation amplitude $d\approx O(10^{-2}[m])$, and
angular frequency $\omega \approx O(10^{3}[\frac{rad}{sec}])$  as a set of
reasonable parameters.

To keep track of order-of-magnitude estimates, we  introduce dimensionless
variables. We define the dimensionless four-vector
$(\tau,\vec{\xi})\,:=\,\frac{1}{d}\,(ct,\vec{x})$ and use the oscillation
amplitude $d$ to represent the length scale. The energy momentum tensor
${\bf T}_{\mu\nu}$ is related  to the dimensionless $T_{\mu \nu}$ by
${\bf T}_{\mu \nu} =: \frac{m c^{2}}{d^{3}}T_{\mu \nu}$, and the Einstein
tensor ${\bf G}_{\mu \nu}$ which has the dimensions $[m^{-2}]$ introduces
${\bf G}_{\mu\nu} =: \frac{1}{d^{2}} G_{\mu \nu}$. The Einstein  equations,
written in terms of dimensionless variables, yield
\begin{equation}
G_{\mu \nu} = \; \frac{8 \pi m G}{c^{2}\; d}\;\; T_{\mu
\nu}\;\;\;,
\label{dimlEe}
\end{equation}
in which $G \approx 6.67 \cdot 10^{-11}[\frac{m^{3}}{sec^{2}kg}]$ is the
universal gravitational constant. The number
$\frac{8 \pi G m}{c^{2} d} \approx 10^{-25}$ estimates the strength of
gravitational signals due to $m$. We bound the mass of $M$ by
$\frac{M}{m}<O(10^{2})$, such that all considerations about $m$ also
apply to $M$. The RHS of Eq. (~\ref{dimlEe}) is very  small if
$\parallel T_{\mu \nu} \parallel\ll 10^{25}$, and the linearized Einstein
equations can be used in this case. Consequently, one expands the
space-time metric $g_{\mu\nu} = \eta_{\mu\nu} + h_{\mu\nu}$ around the
flat Minkowski metric $\eta_{\mu\nu}$, and introducing
$\gamma_{\mu\nu}:=h_{\mu\nu}-\frac{1}{2}\eta_{\mu\nu}h^{\lambda}_{\lambda}$,
Eq. (~\ref{dimlEe}) reduces to
\begin{equation}
\left(\partial^{2}_{\tau} - \Delta_{\vec{\xi}}\right)\;
\gamma_{\mu\nu}(\tau,\vec{\xi}) = \; - \;
\frac{16 \pi m G}{c^{2} d} \; \; T_{\mu\nu}(\tau,\vec{\xi}) \;\;\; ,
\label{linEinstEq}
\end{equation} by use of the Hilbert gauge condition
$\gamma^{\nu}_{\mu,\nu}=0$ $[8]$. $\Delta_{\vec{\xi}}$ is the Laplace
operator in the dimensionless spatial coordinates
$\vec{\xi}$. Eq. (~\ref{linEinstEq}) is solved by the superposition of a
particular solution with a homogenous solution. The full particular
solution includes the influences of both $M$ and $m$. We impose the
boundary condition
$\gamma_{\mu\nu}\rightarrow 0$ in the limit $|\vec{\xi}|\rightarrow
\infty$. Since the full particular solution is asymptotically zero, as
will be shown in section II.2, this requires the homogenous solution,
given by a superposition of plane waves with constant amplitudes, to be
zero. Therefore,
$\gamma_{\mu\nu}$ is the particular solution of Eq. (~\ref{linEinstEq}).
Due to linearity,  the contributions from $M$ and $m$ to $\gamma_{\mu\nu}$
and $T_{\mu\nu}$ are additive and can be discussed separately, thus
$\gamma_{\mu\nu}=\gamma^{(m)}_{\mu\nu}+\gamma^{(M)}_{\mu\nu}$ and
$T_{\mu\nu}=T^{(m)}_{\mu\nu}+T^{(M)}_{\mu\nu}$. We will first analyze the
influences due to $m$, determined by $\gamma^{(m)}_{\mu\nu}$ and
$T^{(m)}_{\mu\nu}$. We obtain
\begin{equation}
\gamma^{(m)}_{\mu\nu}(\tau,\vec{\xi}) = \; - \; \frac{4 m G}{c^{2}d}
\int d \tau' d^{3}\vec{\xi}' \; \frac{T^{(m)}_{\mu\nu} (\tau',\vec{\xi}')}
{|\vec{\xi} -
\vec{\xi}'|}\; \delta(\tau' - \tau + |\vec{\xi} - \vec{\xi}'|) \;
\Theta(\tau - \tau')\;,
\end{equation}
in which $\delta$ is the Dirac-delta distribution and $\Theta$ is the
Heaviside function. The perturbation parameter of our analysis is the
ratio $\beta = \frac{\omega d}{c} \approx O(10^{-8})$ between the maximal
velocity
$\omega d$ of the oscillating mass $m$ and the speed of light $c$. Our
analysis will show that the retardation is an effect of the order
$O(\beta)$. Therefore, a relative accuracy of the order $O(\beta^{2})$ is
sufficient for our calculations. $T^{(m)}_{00}$ can be decomposed into
\begin{eqnarray*}
T^{(m)}_{00}(\tau',\vec{\xi}')\,=\,T_{00}^{(m,0)}(\tau',\vec{\xi}')\,+\,
T_{00}^{(m,2)}(\tau',\vec{\xi}')\;,
\end{eqnarray*}
where $T_{00}^{(m,i)}$ is of order $O(\beta^{i})$. Note
that this decomposition does not correspond to the full Taylor expansion
with respect to $\beta$. $T_{00}^{(m,0)}$ is the  only contribution of
order $O(1\,=\,\beta^{0})$ to $T_{\mu\nu}$, and is determined by the
dimensionless rest energy density $\rho(\tau',\vec{\xi}')$ of $m$, which
satisfies $\int d\vec{\xi}'\rho(\tau',\vec{\xi}')=1$. In all physical
systems,
$\rho(\tau',\vec{\xi}')$ can be assumed to be a sharply-peaked, smooth
function with $\sup_{\vec{\xi}'}\,\rho(\tau',\vec{\xi}')\ll O(10^{25})$,
such that Eq. (~\ref{linEinstEq}) can be used. For mathematical
convenience, we replace
$\rho(\tau',\vec{\xi}')$ by a Dirac delta distribution
$\delta^{(3)}$, and consequently obtain
\begin{equation}
T_{00}^{(m,0)}(\tau',\vec{\xi}') =
\delta^{(3)}(\vec{\xi}' - \hat{z} \sin \beta \tau') \;.
\end{equation}
Note that the space and time dependent terms on the RHS of Eq. (5) cannot
be separated, as opposed to the expression $\rho(\vec{\xi'})\,\exp[i\omega
\tau']$ commonly found in the literature which is suitable for multipole
decomposition.
$T_{00}^{(m,2)}$ is determined by the sum of the kinetic energy of $m$
and the potential energy  of the spring, and is of order $O(\beta^{2})$.
The components $T^{(m)}_{0i}$, $i=1,2,3$, are given by
\begin{eqnarray*}
T^{(m)}_{0i}(\tau',\vec{\xi}') = \beta\;\hat{z}_{i}
\;\cos\beta\tau'\;\delta^{(3)}(\vec{\xi}' - \hat{z} \sin \beta \tau')\;,
\end{eqnarray*}
and are of order $O(\beta)$. The components $T^{(m)}_{ij}$,
$i,j=1,2,3$, given by
\begin{eqnarray*}
T^{(m)}_{ij}(\tau',\vec{\xi}') =
\beta^{2}\;\hat{z}_{i}\;\hat{z}_{j}
\;\cos^{2} \beta
\tau'\;
\delta^{(3)}(\vec{\xi}' - \hat{z} \sin \beta \tau') \;,
\end{eqnarray*}
will only contribute to error terms of the order $O(\beta^{2})$.

We evaluate $\gamma^{(m)}_{\mu\nu}(\tau,\vec{\xi})$ at the point
$\vec{\xi}\,=\,\frac{r}{d}\hat{z}$ on the  positive $z$ axis, such that
$\xi_{3} = \frac{r}{d}\,>\,1$. Inserting Eq. (5) into Eq. (4), we first
calculate $\gamma_{0 0}^{(m,0)}(\tau,\xi_{3}\hat{z})$, defined by
$\gamma^{(m)}_{00}=\gamma_{0 0}^{(m,0)}+\gamma_{0 0}^{(m,2)}$. The
$\delta$-integration in $\xi_{1}$ and $\xi_{2}$ can be readily carried
out, yielding
\begin{eqnarray}
\gamma_{0 0}^{(m,0)}(\tau,\xi_{3}\hat{z}) &=&\; - \;
\frac{4 m G}{c^{2} d}
\int d\xi_{3}' d\tau' \; \frac{\delta(\xi_{3}' - \sin \beta \tau')}
{\xi_{3}-\xi_{3}'} \;
\delta(\tau' + \xi_{3} - \xi_{3}' - \tau) \;
\Theta(\tau - \tau')
\nonumber
\\ &=&  - \; \frac{4 m G}{c^{2} d}
\int d\tau' \; \frac{\delta(\tau' + \xi_{3} - \sin \beta \tau' - \tau)}
{\xi_{3} - \sin \beta \tau'} \;
\Theta(\tau - \tau')\;.
\end{eqnarray}
Further reduction is possible using the formula
\begin{eqnarray*}
\int d\tau' \; \delta\left(f(\tau')\right) \;g(\tau') = \sum_{\theta \in
\lbrace {\theta|f(\theta) = 0} \rbrace} \;
\frac{1}{f'(\theta)}\; \; g(\theta) \; \;,
\end{eqnarray*}
in which  we identify the zeros of the argument in the
delta distribution as the roots of $\theta$ in the equation
\begin{equation}
\theta(\tau,\xi_{3}) = \sin \beta \theta(\tau,\xi_{3}) + \tau - \xi_{3}
\;\;.
\label{theta}
\end{equation}
Note that $\theta(\tau,\xi_{3})$ is a function of the
combination $\tau -
\xi_{3}$. The derivative of the argument of the delta in Eq. (6) yields
$1 -\beta\cos\beta \theta(\tau,\xi_{3})$ at $\theta(\tau,\xi_{3})$.  We
obtain for $\gamma_{00}^{(m,0)} (\tau,\xi_{3}\hat{z})$
\begin{equation}
\gamma_{0 0}^{(m,0)}(\tau,\xi_{3}\hat{z}) = \; - \;
\frac{4 mG}{c^{2} d}\;\;
\frac{1}{1 - \beta \cos \beta \theta(\tau,\xi_{3})} \;\;
\frac{1}{\xi_{3} - \sin \beta\theta(\tau,\xi_{3})} \;.
\label{sol1}
\end{equation}
In Eq. (~\ref{sol1}),
$\beta$ is not only a small perturbation parameter, but  also a
"dimensionless angular frequency". In contrast to $\beta$, the arguments
$\beta\theta(\tau,\xi_{3})$ of the trigonometric functions need not be
small for arbitrary $\tau$ or $\xi_{3}$, and cannot be used as
perturbation parameters. Since $T_{00}^{(m,2)}$ is of order
$O(\beta^{2})$, it follows that $\gamma_{00}^{(m,2)}= O(\beta^{2})$ is an
error term, and that
$\gamma^{(m)}_{00}(\tau,\vec{\xi}\hat{z})=\gamma_{00}^{(m,0)}(\tau,\vec{\xi}
\hat{z})
+O(\beta^{2})$. Hence it is sufficient to analyze $\gamma_{0 0}^{(m,0)}$
to the order of $O(1=\beta^{0})$ and $O(\beta)$, to determine
$\gamma^{(m)}_{0 0}$ with a relative error of order $O(\beta^{2})$. The
solutions for
$\gamma^{(m)}_{0i}(\tau,\xi_{3}\hat{z})$, $i=1,2,3$, can be derived in a
similar manner, yielding
\begin{eqnarray*}
\gamma^{(m)}_{0i}(\tau,\xi_{3}\hat{z}) = \; - \;\frac{4 m G}{c^{2} d}
\;\;
\frac{\hat{z}_{i}\;\beta \cos \beta \theta(\tau,\xi_{3})}{1 - \beta \cos
\beta
\theta(\tau,\xi_{3})} \;\;
\frac{1}{\xi_{3} - \sin \beta\theta(\tau,\xi_{3})} \;.
\end{eqnarray*}
The solutions for
$\gamma^{(m)}_{ij}(\tau,\xi_{3}\hat{z})$, $i,j=1,2,3$, which are given by
\begin{eqnarray*}
\gamma^{(m)}_{ij}(\tau,\xi_{3}\hat{z}) = \; - \;\frac{4 m G}{c^{2} d}
\;\;
\frac{\hat{z}_{i}\hat{z}_{j}\;\beta^{2} \cos^{2} \beta
\theta(\tau,\xi_{3})} {1 -\beta
\cos \beta
\theta(\tau,\xi_{3})} \;\;
\frac{1}{\xi_{3} - \sin \beta\theta(\tau,\xi_{3})} \;,
\end{eqnarray*}
contribute to error terms of order $O(\beta^{2})$. The contributions
$\gamma^{(M)}_{\mu\nu}(\tau,\xi_{3}\hat{z})$ of the mass $M$ can be
obtained by substituting $d\rightarrow D$,
$r\rightarrow R$, $m\rightarrow M$,
$\beta\rightarrow -\frac{D}{d}\beta$, $\theta\rightarrow
\frac{d}{D}\theta$,
$(\tau,\vec{\xi})\rightarrow \frac{d}{D}(\tau,\vec{\xi})$,
$\xi_{3}=\frac{r}{d}\rightarrow \frac{R}{D}$ in
$\gamma^{(m)}_{\mu\nu}(\tau,\xi_{3}\hat{z})$, where $D=\frac{m}{M}d$ is
the oscillation amplitude of $M$, and $R$ is its average distance to the
point of measurement. This is because both masses oscillate sinusoidally
with the same frequency $\omega$ and opposite phases along the $z$ axis.

It follows that $\partial_{\tau}\gamma^{(m,M)}_{0i}(\tau,\xi_{3}\hat{z})$
is of  order $O(\beta^{2})$ for both $M$ and $m$. We will show below that
the only Levi-Civita connection coefficients needed for our analysis are
\begin{eqnarray}
\Gamma^{\mu}_{00}(\tau,\xi_{3}\hat{z})&=&
-\frac{1}{2}(2h_{0\mu,0}(\tau,\xi_{3}\hat{z})
-h_{00,\mu}(\tau,\xi_{3}\hat{z}))
\nonumber
\\ &=&
\frac{1}{4}\,\gamma_{00,\mu}(\tau,\xi_{3}\hat{z})
-\gamma_{0\mu,0}(\tau,\xi_{3}\hat{z})
+\frac{1}{4}\,\gamma^{k}_{k,\mu}(\tau,\xi_{3}\hat{z})\;.
\label{LeviCivita}
\end{eqnarray}
Thus, $\Gamma^{i}_{00}=\frac{1}{4}\,\gamma_{00,i}^{(0)}+O(\beta^{2})$ for
$i=1,2,3$. Here, we have used that
$h_{\mu\nu}=\gamma_{\mu\nu}-\frac{1}{2}\eta_{\mu\nu}
\gamma^{\lambda}_{\lambda}$ and
$\Gamma^{\mu}_{\rho\sigma}=\frac{1}{2}\eta^{\mu\nu}(h_{\rho\nu,\sigma}+
h_{\sigma\nu,\rho}-h_{\sigma\rho,\nu})$, which include the influences of
both $M$ and $m$.

Given the metric $g_{\mu\nu}=\eta_{\mu\nu} + h_{\mu\nu}$, we intend to
study the motion of a test mass point
$\tilde{m}\ll m$ in the immediate vicinity of $m$, that is part of
another  mechanical two-body oscillator on the
$z$ axis. The dimensionless space-time coordinates of $\tilde{m}$ are
scaled with $d$, yielding
$(\xi^{0}(\tau),\vec{\xi}(\tau))$. We are only interested in oscillatory
solutions with small amplitudes of order
$\approx d\cdot O(\beta)$, which are induced by the oscillations of the
space-time metric.  Therefore, we require that the dimensionless velocity
components $\xi_{,0}^{i}$,
$i=1,2,3$, of $\tilde{m}$, are of order
$O(\beta^{2})$. The subscript "{\scriptsize ,0}" denotes the derivative
with respect to $\tau$. The proper time of the test mass satisfies
$\xi^{0}(\tau)\approx\tau$ and $\xi_{,0}^{0}=O(1)$, since the deviations
from flat Minkowskian space-time induced by $M$ and $m$ are very small.
The action which  determines the dynamics of the mass
$\tilde{m}$ in our coordinate system is given by
\begin{eqnarray*}
S= \tilde{m}c d \int_{\tau_{0}}^{\tau_{1}} d\tau
\left\lbrace-\sqrt{g_{\mu\nu}\xi_{,0}^{\mu}\xi_{,0}^{\nu}}- U(\xi^{i})
\right\rbrace\;,
\end{eqnarray*}
where $U(\xi^{i})$ is the dimensionless potential energy
due to the spring. For the moment, we assume that the second mass of the
pick-up system is "infinitely" larger than
$\tilde{m}$, and static.  The space-time trajectory of the test mass
$\tilde{m}$ is determined by the Euler-Lagrange equation
\begin{eqnarray*}
-\frac{\xi_{,00}^{\mu}+
\Gamma^{\mu}_{\rho\sigma}\xi_{,0}^{\rho}\xi_{,0}^{\sigma}}
{\sqrt{g_{\mu\nu}\xi_{,0}^{\mu}\xi_{,0}^{\nu}}} + U_{,\mu} =0\;,
\end{eqnarray*}
where $\Gamma^{\mu}_{\rho\sigma}$ are linearized Levi-Civita connection
coefficients. The dominant terms in the sum
$\Gamma^{\mu}_{\rho\sigma}\xi_{,0}^{\rho}\xi_{,0}^{\sigma}$ arise from
$\Gamma^{\mu}_{00}\xi_{,0}^{0}\xi_{,0}^{0}$ . All terms
$\Gamma^{i}_{\rho\sigma}\xi_{,0}^{\rho}\xi_{,0}^{\sigma}$ which involve
$\xi_{,0}^{j}$, $j=1,2,3$, are at least a factor of order
$O(\beta^{2})$ smaller, and can be neglected. From the equation of motion
for
$\xi^{0}$ follows that $\frac{\xi_{,00}^{0}}{(\xi_{,0}^{0})^{2}} +
\Gamma^{0}_{00} =0$ with a relative error of order $O(\beta^{2})$. Using
the fact that
$\Gamma^{0}_{00}$ is a total derivative with respect to $\tau$, and that
$\parallel\gamma_{\mu\nu}\parallel\ll O(\beta^{2})$, one  obtains
$\xi_{,0}^{0} = 1 + O(\beta^{2})$ and
$\sqrt{g_{\mu\nu}\xi_{,0}^{\mu}\xi_{,0}^{\nu}} = 1 + O(\beta^{2})$.
Consequently, the motion of $\tilde{m}$ is determined by
\begin{eqnarray}
\xi_{i,00} + U_{,i} +
\frac{1}{4}\,\gamma_{00,i}^{(0)}=0\;,
\label{EuLa}
\end{eqnarray}
with a relative error of the order $O(\beta^{2})$.  The
contributions of $M$ and $m$ to the "gravitational force"
$\frac{1}{4}\,\gamma_{00,i}^{(0)}$ are additive, and have the same
functional form. This weak-curvature limit is usually referred to as the
post-Newtonian approximation, or the Newtonian limit. The pick-up
oscillator is typically driven at resonance, but has a finite $Q$ factor.
In all realistic systems, the condition that the velocity is of order
$\approx O(\beta^{2})$ is satisfied, because the gravitational excitation
of the pick-up system is very weak, and  Eq. (~\ref{EuLa}) can be used.

The Levi-Civita connection coefficients Eq. (~\ref{LeviCivita}) depend on
our specific choice of coordinates, or gauge. It is not a priori clear
that these coefficients are the correct expression for the physical
gravitational force, since they could be influenced by meaningless
coordinate effects. The physically measurable force is determined by the
change of these coefficients in the absence of the source, i.e. for
$T_{\mu\nu}=0$, calculated in the same gauge, and using the same boundary
conditions. In the Hilbert gauge,
$T_{\mu\nu}=0$ results in the vacuum field equations
$\left(\partial^{2}_{\tau}-\Delta_{\vec{\xi}}\right)\,
\gamma_{\mu\nu}(\tau,\vec{\xi}) = \,0$. The solution is a superposition of
plane waves with constant amplitudes, and must be zero due to the boundary
condition $\gamma_{\mu\nu}\rightarrow 0$ for
$|\vec{\xi}|\rightarrow \infty$. Consequently, the corresponding
Levi-Civita connection coefficients are zero, which shows that the terms
$\Gamma^{i}_{00}$ in Eq. (~\ref{LeviCivita}), $i=1,2,3$, correctly
describe the physical force.

In the Newtonian limit, $\frac{c^{2}}{4}$ times (8) is interpreted as the
classical gravitational potential
\begin{equation}
V(\tau,\xi_{3}\hat{z}) = \; - \;\frac{m G}{d} \;\;
\frac{1}{1 - \beta \cos \beta \theta(\tau,\xi_{3})} \;\;
\frac{1}{\xi_{3} - \sin \beta\theta(\tau,\xi_{3})} \;\;
\label{gravpot}
\end{equation}
generated by $m$, which acts on the external mass point
$\tilde{m}$ by a gravitational force that is proportional to the
gradient  of
$V$, according to Eq. (~\ref{EuLa}). The physical meaning of the solution
Eq. (~\ref{gravpot}) becomes more  apparent if one expands the first
fraction in a geometric series, and the second fraction in a Fourier
series
\begin{eqnarray}
\lefteqn{V(t,r \hat{z}) = \; \;
 - \; \frac{m G}{r} \;\;
\left\lbrace 1 + \beta \cos \beta \theta(\frac{c}{d}\;t\;,\;\frac{r}{d})
+O(\beta^{2})\right\rbrace \,\cdot\;}\;
\nonumber
\\ & &\left\lbrace \;\frac{1}{\sqrt{1-\left(\frac{d}{r}\right)^{2}}} \, +
\,
\left(\frac{r}{d}\right)\left[\,
\frac{2}{\sqrt{1-\left(\frac{d}{r}\right)^{2}}}\,-\,2 \,\right] \,
\sin \,\beta \,\theta(\frac{c}{d}\,t\,,\,\frac{r}{d}) \,+ \right.
\nonumber
\\ & &
\sum_{n>2,\,n\,odd}\;
\left(\frac{d}{r}\right)^{n}\;\;2^{1-n}\;\;\left[\;\;1 + c_{n}
\;\;\right]
\; \sin n \;\beta\; \theta(\frac{c}{d}\;t\;,\;\frac{r}{d}) \,+
\;\;\;\;
\nonumber
\\ & & \left. \sum_{n>1,\,n\,even}
\left(\frac{d}{r}\right)^{n}\;\;2^{1-n}\;\;\left[\;\;1 + d_{n}
\;\;\right]
\; \cos n \;\beta\; \theta(\frac{c}{d}\;t\;,\;\frac{r}{d})
\;\;\;\;
\right\rbrace \;\;\; ,
\label{potFour}
\end{eqnarray}
in which all coefficients $c_{n}$ and $d_{n}$ are of order
$O((\frac{d}{r})^{2})$. The leading terms of order
$O(1=\beta^{0})$ can be  obtained by setting $\beta$ equal to zero in the
first bracket. The time-independent contribution
$\frac{ m G}{r}\lbrace1-\left(\frac{d}{r}\right)^{2}\rbrace^{\,-1/2}$  in
the Fourier expansion is the static potential of a resting mass
distribution centered at the origin,  and converges to the classical
Newtonian potential
$\frac{m G}{r}$ for large  $r$. The next, first harmonic  term has the
asymptotic form
$\frac{m G d}{r^{2}}(1+O((\frac{d}{r})^{2}))\sin\beta\theta(\frac{c}{d}t,
\frac{r}{d})$. For large $r$, it describes a propagating, sinusoidal
oscillation  proportional to $\frac{1}{r^{2}}$, which decays more rapidly
than the usual radiative $\frac{1}{r}$ solutions. All following terms are
higher harmonics  that decay even faster with increasing distance. The
contributions of
$O(\beta)$ exhibit the same asymptotic behaviour with respect to
$\frac{d}{r}$.

\subsection{\normalsize \bf GRAVITATIONAL EFFECTS OF THE MASS $M$}

In this section, we will discuss the gravitational effects of the mass
$M$ in the immediate vicinity of
$m$. The average distance $R$ between $M$ and the point of measurement,
lying in the immediate vicinity of $m$, is much larger than $d$, and also
much larger than $r$. Due to $M\gg m$, the oscillation amplitude
$D =\frac{m}{M}d$ of $M$, which is calculated from classical mechanics,
is much smaller than
$d$. Both masses oscillate at frequency $\omega$ along the $z$ axis, and
the gravitational potential due to $M$  at
$\xi =\frac{r}{d}\hat{z}$ can be obtained by inserting $d\rightarrow D$,
$r\rightarrow R$, $m\rightarrow M$,
$\beta\rightarrow -\frac{D}{d}\beta$, $\theta\rightarrow
\frac{d}{D}\theta$,
$(\tau,\vec{\xi})\rightarrow \frac{d}{D}(\tau,\vec{\xi})$,
$\xi_{3}=\frac{r}{d}\rightarrow \frac{R}{D}$ in Eq. (~\ref{potFour}). The
static term is approximately
$-\,\frac{MG}{R}$ and can dominate over the static field stemming from
$m$. However, one obtains approximately $\frac{MG}{R}\frac{D}{R}\,=\,
\frac{mGd}{r^{2}}(\frac{r}{R})^{2}$ for the amplitude of the first
time-dependent term in Eq. (~\ref{potFour}), which is a factor of
$(\frac{r}{R})^{2}$ {\em smaller} than the corresponding term due to $m$.
The amplitude of the first harmonic  term due to $M$ has the {\em
opposite sign} as  compared to corresponding term due to $m$, because $M$
and $m$ oscillate with opposite phases.

In addition, the amplitudes of the higher harmonic terms, approximately
given by
$\frac{MG}{R}(\frac{D}{R})^{n}\,=\,\frac{mG}{r}(\frac{d}{r})^{n}
(\frac{r}{R})^{n+1}(\frac{m}{M})^{n-1}$, are a factor of
$(\frac{r}{R})^{n+1}(\frac{m}{M})^{n-1}$ smaller than the corresponding
terms due to $m$. One can therefore neglect the  {\em time-dependent}
gravitational influences due to $M$ if $\frac{r}{R}$ is small, which is
the case in the immediate  vicinity of
$m$. At larger distances from $m$, the  number $\frac{r}{R}$ approaches
$1$, and all radiative effects cancel,  which is required by momentum
conservation.

\subsection{\normalsize \bf RETARDATION OF GRAVITATIONAL FORCES}

It will first be shown that the forces derived from the gravitational
potential Eq. (~\ref{gravpot}) are not retarded in the region known as the
"near zone" in the literature, which is in accordance to
$[7]$.  Then, it will be proved that these gravitational forces are
retarded in the immediate vicinity of $m$. As discussed in the derivation
of Eq. (~\ref{EuLa}), we place a small test mass
$\tilde{m}$ which is part of a resonant mechanical oscillator at position
$\vec{\xi}\,=\,r\,\hat{z}$ on the $z$ axis to detect the gravitational
forces generated by the source $m$. From
$F(\tau,\frac{r}{d}\hat{z})=-\,\tilde{m}\,\partial_{r}\,V(\tau,\frac{r}{d}
\hat{z})$ it follows that the gravitational force acting on $\tilde{m}$ is
\begin{equation}
F(\tau,\frac{r}{d}\hat{z}) = \; - \;\frac{\tilde{m} m
G}{d^{2}}\;
\left\lbrace\left(\frac{1}{1 - \beta \cos \beta \theta(\tau,\frac{r}{d})}
\;
\frac{1}{\frac{r}{d} - \sin \beta\theta(\tau,\frac{r}{d})}\right)^{2}
\;+\;O(\beta^{2})\right\rbrace\;.
\label{force1}
\end{equation}
$F(\tau,\frac{r}{d}\hat{z})$ points along the $z$ axis, whereas the other
force components vanish due to symmetry reasons. Eq. (~\ref{force1}) is
accurate to  orders
$O(1=\beta^{0})$ and $O(\beta)$, but not to the order $O(\beta^{2})$,
since
$T_{00}^{(m,2)}$  has not been considered in
$\gamma_{00}^{(m,0)}$. For our analysis it suffices that the results
derived from  Eq. (~\ref{force1}) are conclusive to $O(1=\beta^{0})$ and
$O(\beta)$. To obtain Eq. (~\ref{force1}),  we have used the fact that
the  derivative of
$\theta(\tau,\frac{r}{d})$ with respect to $r$ is given by
\begin{equation}
\partial_{r}\;\theta(\tau,\frac{r}{d})\;=\;-\;\frac{1}{d}\;\;
\frac{1}{1\;-\;\beta\,\cos\;\beta\;\theta(\tau,\frac{r}{d})}
\label{dertheta}
\end{equation}
which follows from
\begin{eqnarray}
\partial_{r}\,\theta(\tau,\frac{r}{d})\;&=&\;\partial_{r}\;\sin\;\beta\;
\theta(\tau,\frac{r}{d})\;-\;\frac{1}{d}
\nonumber
\\ &=&\;\beta\;\cos\;\beta\;\theta(\tau,\frac{r}{d})\;\;
\partial_{r}\,\theta(\tau,\frac{r}{d})
\,-\,\frac{1}{d}\;\;.
\end{eqnarray}
Note that the leading term in $F(\tau,\frac{r}{d})$ is proportional to
$V(\tau,\frac{r}{d}\hat{z})^{\,2}$. The expression for the gravitational
force generated by
$M$ is similar to Eq. (~\ref{force1}). It has a static component  which is
much stronger than the static term due to $m$. However, the time-dependent
gravitational forces generated by $M$  are negligible in the immediate
vicinity of $m$, because
$\frac{d}{r}\,\approx\,1$ and therefore $\frac{r}{R}\,\ll\,1$. At large
distances from
$m$, $M$ and $m$ produce time-dependent gravitational forces of the same
strength, since $\frac{d}{r}\,\rightarrow\,0$ and
$\frac{r}{R}\,\rightarrow\,1$.

\subsubsection{GRAVITATIONAL FORCES IN THE "NEAR ZONE"}

The spherical region $\lbrace r=|\vec{x}|;\beta \ll\frac{d}{r} \ll
1\rbrace$ is commonly defined as the "near zone"  in the literature. For
this region, the approximation
$\theta(\tau,\frac{r}{d})\,\approx\,\tau\,-\,\frac{r}{d}$ can be used,
which is familiar from radiation theory. In addition, we have used the
identity
$\frac{1}{1+y}= 1-y +O(|y|^{2})$, with $|y| < 1$, to simplify the
fractions in
$V(\tau,\frac{r}{d}\hat{z})$, and to  reduce  the potential to
\begin{equation}
V(\tau,\frac{r}{d}\hat{z}) \,= \, - \,\frac{m G}{d} \,
\left\lbrace 1 + \beta \cos \beta (\tau\,-\,\frac{r}{d})\right\rbrace \,
\left\lbrace\frac{d}{r} + \left(\frac{d}{r}\right)^{2}
\sin \beta(\tau\,-\,\frac{r}{d})\right\rbrace \,+\,O(\beta^{2})\;.
\label{gravpot1}
\end{equation}
The trigonometric functions can be approximated by their Taylor
expansions to first order in the variable
$\beta\,\frac{r}{d}\;\ll\,1$. Keeping terms only to the orders $O(\beta)$
and
$O(\frac{d}{r})$, one obtains
\begin{eqnarray}
F(\tau,\frac{r}{d}\hat{z}) &=& \, - \,\frac{\tilde{m} m
G}{r^{2}} \,
\left\lbrace 1 + 2\beta \cos \beta\tau\right\rbrace \,
\left\lbrace 1 + 2\frac{d}{r}\sin \beta\tau - 2\beta\,\cos
\beta\tau\right\rbrace
\,+\,O(\beta^{2})
\nonumber
\\ &=&\, - \,\frac{\tilde{m} m G}{r^{2}} \,
\left\lbrace 1 + 2\frac{d}{r}\sin \beta\tau
 + 4\beta\, \frac{d}{r}\,\cos \beta\tau \,\sin
\beta\tau\right\rbrace\,+\,O(\beta^{2})
\label{gravpot2}
\end{eqnarray}
for the gravitational force acting on the test mass $\tilde{m}$. The
term of order $O(\beta\, \frac{d}{r})\ll O(\beta)$ is a negligible
contribution to the second  harmonics. Hence the contributions of order
$O(\beta)$  {\em cancel}. The time-dependent term of order
$O(1=\beta^{0})$ in the force acts  instantaneously, thus there  is no
measurable retardation in the "near zone" where
$\frac{d}{r}\,\ll\,1$. The electromagnetic version of this result can be
found in $[7]$. Historically, this result has convinced the physics
community that there exists no region at all where any retardation can be
observed.

The additional  constraint $\frac{r}{R}\,\ll\,1$ allows the influence of
the second mass $M$ to be neglected. As $r$ increases, this condition can
not be satisfied, and $\frac{r}{R}\,\rightarrow\,1$. The complete
gravitational potential in the "near zone" limit, which includes the
contributions of both
$m$ and $M$ is given by
\begin{eqnarray*}
V_{tot}(\tau,\frac{r}{d}\hat{z})\;
\approx\;-\frac{mG}{r}\left\lbrace 1 +
\frac{M}{m}\;\frac{r}{R}\right\rbrace\;
-\;\frac{mGd}{r^{2}}\left\lbrace\sin\,\beta\,\tau\;-\;\left(\frac{r}{R}
\right)^{2}
\sin\,\beta\,\tau\right\rbrace\;\;,
\end{eqnarray*}
which follows from the discussion in section II.1. The
time-dependent part of this solution vanishes in the limit
$\frac{r}{R}\,\rightarrow\,1$, and there remains a static solution which
is dominated  by the larger mass $M$. Note that $\frac{(m+M)G}{r}$ is  the
rotation symmetric limit solution for $r\rightarrow \infty$, which
corresponds to the asymptotically vanishing particular solution of Eq.
(~\ref{linEinstEq}). We conclude that our result is consistent with
$[7]$, and that it is in accordance with the conservation of momentum.

\subsubsection{GRAVITATIONAL FORCES IN THE IMMEDIATE VICINITY OF THE
SOURCE}

The immediate vicinity of the source is characterized by
$\frac{d}{r}\,\approx\,1$,  in contrast to $\frac{d}{r}\,\ll\,1$ which is
a  necessary condition for radiation theory. Consequently, radiation
theory breaks down. At this distance  range,
$\frac{r}{R}\,\ll\,1$ holds, and time-dependent influences of $M$ are
negligible, as shown in section II.1. As opposed to the situation in the
"near zone", it is not possible to approximate functions of $\frac{d}{r}$
by the leading terms in their Taylor series. For this reason,  we will
have to explicitly calculate the dominant Fourier components of the
lowest harmonics  in Eq. (~\ref{force1}). To order $O(\beta)$, we find
\begin{eqnarray}
F(\tau,\frac{r}{d} \hat{z})  &=&
 - \; \frac{\tilde{m} m G}{d^{2}} \;\;
\left( 1 + 2\beta \cos \beta \theta(\tau\,,\,\frac{r}{d})\right)
\left(\frac{1}{\frac{r}{d} - \sin
\beta\theta(\tau,\frac{r}{d})}\right)^{2}
\; +\;O(\beta^{2})\;.
\label{potFourlead}
\end{eqnarray}
The condition $\beta\,\ll\,\frac{d}{r}$ which has defined
the "near zone"  is also valid for the immediate neighborhood of the
oscillating mass point. Therefore, it is still possible to expand the
trigonometric functions with respect to $\beta\,\frac{r}{d}\approx
O(\beta)$. From the defining  equation Eq. (~\ref{theta}) for
$\theta(\tau,\frac{r}{d})$, it can be seen that
$\beta\,\theta(\tau,\frac{r}{d})$ is a function of the variable
$\beta\,\frac{r}{d}$. From  Eq. (~\ref{dertheta}), one finds
\begin{equation}
\beta\;\theta(\tau,\;\frac{r}{d})\;=\;\beta\;\theta(\tau,0)\;-\;
\beta\;\frac{r}{d}
\;+\;O\left(\beta^{2}\right)
\label{thetaTaylor}
\end{equation}
for the leading terms of the Taylor series in
$\beta\,\frac{r}{d}$. We expand the trigonometric functions in Eq.
(~\ref{potFourlead}) with respect to
$\beta\;\theta(\tau,\frac{r}{d})\,\approx\,\beta\,\theta(\tau,0)\,-\,
\beta\,\frac{r}{d}$, and perform the Taylor series to first order in
$\beta$ in the second bracket. Consequently, the RHS of Eq.
(~\ref{potFourlead}) yields
\begin{eqnarray}
 - \, \frac{\tilde{m} m G}{r^{2}} \,
\left( 1 + 2\beta \cos \beta \theta(\tau\,,\,0)\right)\,
\left(\frac{1}{1 - \frac{d}{r}\sin \beta\theta(\tau,0)}
\,-\,\frac{\beta\,\cos\beta\theta(\tau,0)} {\left(1 - \frac{d}{r}\sin
\beta\theta(\tau,0)\right)^{2}}\right)^{2}
\; +\;O(\beta^{2})\;.
\nonumber
\end{eqnarray}
The expansion of this product to order $O(\beta)$ gives
\begin{eqnarray}
 - \, \frac{\tilde{m} m G}{r^{2}} \,
\left(\frac{1}{\left(1 - \frac{d}{r}\sin \beta\theta(\tau,0)\right)^{2}}
\,-\,\frac{\frac{d}{r}\beta\,\sin 2\beta\theta(\tau,0)} {\left(1 -
\frac{d}{r}\sin \beta\theta(\tau,0)\right)^{3}}\right)\,+
\,O(\beta^{2})\;\;.
\nonumber
\end{eqnarray}
Calculation of the Fourier components of the first harmonics to order
$O(\beta)$ results in
\begin{eqnarray} - \, \frac{\tilde{m} m G}{r^{2}} \,
\frac{1}
{\left(1\,-\,\left(\frac{d}{r}\right)^{2}\right)^{\frac{3}{2}}}\left\lbrace
1\,+\,  2\,\frac{d}{r}
\sin \beta\theta(\tau,0)\,-\,2\beta\Phi(\frac{d}{r})\cos\beta
\theta(\tau,0)\right\rbrace
 +\,h.h.\,+\,O(\beta^{2})\,,
\label{retpot}
\end{eqnarray} where the abbreviation $h.h.$ stands for "higher
harmonics". The function
$\Phi\left( x\right)$ is defined by
\begin{equation}
\Phi\left( x \right)\;=\;\frac{2}{x^{2}}\,
\left(\left(1-x^{2}\right)^{\frac{3}{2}}\;-\;1\;+\;\frac{3}{2}\,x^{2}\right)
\end{equation} for $x\,=\,\frac{d}{r}\,\in\,[0,1]$. $\Phi\left( x\right)$
is a monotonically  increasing function on the unit interval with values
$\Phi(0)\,=\,0$ and
$\Phi(1)\,=\,1$ at the boundaries of its domain, see Fig. 1.
 We have only calculated the first harmonic terms of maximal order
$O(\beta)$ in Eq. (~\ref{retpot}), because we assume that the measurement
in our proposed  experiment is only sensitive to the first harmonics.
Using the relationship
\begin{equation}
\beta \theta(\tau,\Phi(\frac{d}{r})\frac{r}{d})\,=\,\beta\theta(\tau,0)\,
-\,\beta\Phi(\frac{d}{r})\frac{r}{d}\,+\,O(\beta^{2})\;,
\end{equation} which is a generalization of Eq. (~\ref{thetaTaylor}), we
finally obtain
\begin{eqnarray} F(\tau,\frac{r}{d} \hat{z}) \, =\,
 - \, \frac{\tilde{m} m G}{r^{2}} \,
\frac{1}
{\left(1\,-\,\left(\frac{d}{r}\right)^{2}\right)^{\frac{3}{2}}}\left\lbrace
1\,+\,  2\,\frac{d}{r}
\sin \beta\theta(\tau,
\Phi(\frac{d}{r})\frac{r}{d})\right\rbrace \,+
\, h.h.\,+\,O(\beta^{2}) \;.
\label{sol}
\end{eqnarray}
This result shows that the first harmonic  term in the gravitational
force of order $O(\beta)$ can be absorbed into the contribution of order
$O(1\,=\,\beta^{0})$. Moreover, it is obvious from  Eq. (~\ref{theta})
that
$\theta(\tau,\Phi(\frac{d}{r})\frac{r}{d})$ is  a function of the
combination
$\tau-\Phi(\frac{d}{r})\frac{r}{d}$, which shows that the gravitational
force is retarded. For two points
$r_{1}\hat{z}$ and
$r_{2}\hat{z}$ on the oscillator axis, the  relative phase difference
$\Delta\phi$ of the gravitational force is given by
\begin{eqnarray}
\Delta\phi\,&=&\,\beta\,(\Phi(\frac{d}{r_{1}})\frac{r_{1}}{d}
\,-\,\Phi(\frac{d}{r_{2}})\frac{r_{2}}{d})
\nonumber
\\ &=&\,\frac{\omega}{c}\,(\Phi(\frac{d}{r_{1}})\,r_{1}
\,-\,\Phi(\frac{d}{r_{2}})\,r_{2})\;\;,
\label{phsh}
\end{eqnarray}
and is of order $O(\beta)$. At very small distances $r$ from the source,
where
$\frac{d}{r}\,\rightarrow\,1$ and $\Phi(\frac{d}{r})\,\rightarrow\,1$, the
value of the phase shift approximates the result which would occur for a
wave  that travels at speed of light. At larger distances in the "near
zone", we have
$\frac{d}{r}\,\rightarrow\,0$ and
$\frac{r}{d}\Phi(\frac{d}{r})\,\rightarrow\,0$.  Hence there is no
measurable phase shift, and the gravitational forces seem to act
instantaneously, as was shown above. In the intermediate region where
$\frac{d}{r}\,\approx\,1$, the phase shift can be calculated from Eq.
(~\ref{phsh}). Note that in the limit $\frac{d}{r}\rightarrow 0$,
(~\ref{sol}) converges to (~\ref{gravpot2}), as it should be. The absolute
phase shift
$\phi(r):=\frac{\omega}{c}\Phi(\frac{d}{r})r$ converges rapidly to zero
for
$r\rightarrow \infty$, see Fig. 2.

We conclude that there is a possibly measurable retardation of the
gravitational forces in the immediate vicinity of the oscillating source
in our proposed system. A discussion of extended gravitating sources with
the linearized Einstein equations would be a straightforward
generalization of the presented analysis. Due to the superposition
principle, expressions similar to Eq. (~\ref{sol}) would then have to be
integrated over the  spatial spread of the oscillating continuum.
However, we have assumed cylindrical symmetry in our mathematical model,
and thus could find a solution for the dominating gravitational force
term. For systems without cylindrical  symmetry, the analysis will be
much more difficult.  From the results in this section, we will now
discuss how they could be experimentally verified.

\section{\large\bf  EXPERIMENTAL POSSIBILITIES}

Eq. (~\ref{sol}) suggests that it may be possible to experimentally verify
the retardation of gravitational forces in the  immediate vicinity of an
oscillating mass distribution. This would  require both the amplitude and
phase detection of gravitationally transmitted  vibrations. With the
previous mathematical  model in mind, we propose the following experiment:

A mass $m$ within a mass-balanced apparatus is oscillated with a very
stable phase and frequency
$\omega$,  as described in the mathematical model. If a mechanical
resonator with eigenfrequency $\omega$, mass
$\tilde{m}$, and quality factor $Q$ is placed at position $r \hat{z}$,  it
will be driven into resonance by the first harmonic term in the
gravitational  force. Given suitably chosen parameters, the amplitude of
the sinusoidally  vibrating resonator, typically of the order
$O(10^{-9}[m])$, can be detected with modern technology $[5]$.

We propose to measure the retardation of the gravitational force in the
following way. Place the mechanical resonator at the location
$r_{1}\hat{z}$ on the oscillator axis, such that
$\frac{d}{r_{1}}\,\approx\,1$. From the observed gravitationally induced
sinusoidal vibration, the phase difference
$\Delta
\phi_{1}$ between the phase of the resonator  and the phase of the
vibrating source mass can be measured. If the mechanical  resonator is
placed at another location
$r_{2} \hat{z}$ on the oscillator axis, a different result $\Delta
\phi_{2}$ will be observed. By subtracting
$\Delta \phi_{1}$ from $\Delta \phi_{2}$, all internal influences from
the  experimental measuring instruments can be eliminated, resulting in
the relative  phase shift $\Delta \phi_{1}\,-\,\Delta \phi_{2}$. The
value of $c$ in Eq. (~\ref{phsh}) can then be calculated from the
measured relative phase shift
$\Delta \phi_{2} - \Delta \phi_{1}$, which would be the  speed of light
according to General Relativity. The expected values for the relative
phase shift are shown in Fig. 3. Since linearized General Relativity is
formally a special relativistic field theory, measurement of
$c$ with the proposed system would provide a new test of its Lorentz
covariance. Moreover, our proposed laboratory experiment would provide the
first direct observation of propagating gravitational fields. In
addition,  the scalar component of order $O(1=\beta^{0})$ in
Eq. (~\ref{linEinstEq}) also determines the Coulomb potential of an
oscillating point charge after a substitution of parameters. Therefore,
our solution Eq. (~\ref{sol}) also applies to the retarded Coulomb force
generated by an oscillating point charge. Measurement of the retardation in
the immediate vicinity of the charge would thus be another interesting
experimental possibility.

Since the 1960s, several researchers have detected the amplitudes of
gravitationally  induced vibrations in laboratory experiments
$[5,9,10,11]$. It may be possible that these experiments can be analyzed
in a similar way as the model in this paper, which would provide
information about  the magnitude of the retardation, if present. The
detection of the retardation of  gravitational interaction in laboratory
experiments has not been discussed  much recently because the usual "near
zone" calculations predict instantaneous  interactions, as was shown in
section II. However, this result does not  apply in the immediate
vicinity of the  oscillating source, as was also shown in section II.
Note that our experiment is not related to the gravitational waves of
helicity two, which are far more  difficult to detect, since they are of
order
$O(\beta^{2})$.  Unfortunately, several technical problems must be solved
before measurements to test these retardations can be realized. In 1958,
Q. Kerns demonstrated  an analog electronic circuit capable of  measuring
$10^{-6}$ degree phase shifts $[4]$. We have made several modifications to
Kerns' original design, and we are currently able to measure
$10^{-7}$ degree relative phase shifts over a $30$ second measurement
period $[13]$. Currently, the technology is thermally limited, but thermal
control of the system and application of noise reduction techniques may
soon enable us tho increase the phase sensitivity to the
$10^{-9}$ degree phase accuracy required, to measure the retardation to
$1\%$
$[5]$. It is not yet known if the phase stability of the mechanical
systems used for the experiments can be maintained to this accuracy over
the measurement time.

\section{\large \bf CONCLUSIONS}

Using the linearized Einstein  equations, we have proved that the
gravitational force generated by a specific oscillating mass distribution
is retarded in its immediate vicinity.  We have proposed an experiment to
measure this retardation which may enable the first direct experimental
observation of propagating gravitational fields.  The amplitudes of
gravitationally induced vibrations in systems similar to  ours have been
successfully  detected since the 1960s. The detection of the retardation
requires a very sensitive phase measurement, which may now be possible
given the state of modern technology.
\\
\\ {\bf REFERENCES}

\newcounter{ct}
\begin{list}{[\arabic{ct}]}{\usecounter{ct}}

\item J. Cook, "An analysis of methods for measuring the velocity of
gravitational  disturbances", thesis on file at Pennsylvania State
University (1947).

\item J. Cook, "On measuring the phase velocity of an oscillating
gravitational field", J. of the Franklin Inst., Vol. 273, No. 6, June
(1962).

\item I. Bershtein and M. Gertsenshtein, "Possibility of  measuring the
velocity  of propagation of gravitation in the laboratory", J. Expt.
Theort. Phys. U.S.S.R.,  37, pp. 1832-1833 (1959).

\item Q. Kerns, "Proposed laboratory measurement of the  propagation
velocity of  gravitational interaction", UCRL - 8438, Berkeley, CA (1958).

\item W. D. Walker and J. Dual, "Experiment to measure the propagation
speed of  gravitational interaction", Talk presented at Virgo
International Conference on  Gravitational Waves: Sources and  Detectors,
March (1996). To be published in Virgo conference proceedings.

\item W. Misner, K. Thorne, J. Wheeler, "Gravitation", W. H. Freeman and
Company New York, Ch. 36.1 - 36.3,  (1973).

\item R. P. Feynman, "Feynman Lectures in Physics", Vol. 2, Addison
Wesley,   Ch. 21, (1989)

\item N. Straumann, "General Relativity and Relativistic Astrophysics",
texts  and monographs in physics, Springer Verlag Berlin Heidelberg, pp.
218-219 (1991).

\item J. Sinsky and J. Weber, "New source for dynamical gravitational
fields",  Physical Review Letters, Vol. 18, No. 19 (1967).

\item Y. Ogawa, K. Tsubono, H. Hirakawa, "Experimental test of the law of
gravitation", Phys. Rev. D, Vol. 26, No.4, August (1982).

\item  P. Astone et al., "Evaluation and preliminary measurement of the
interaction  of a dynamical gravitational near field with a cryogenic
gravitational wave antenna",  Z. Phys. C 50, pp. 21-29 (1991).

\item Y. T. Chen, A. Cook, "Gravitational Experiments in the Laboratory",
Cambridge University Press, pp. 187-195 (1993).

\item W. D. Walker and J. Dual, "Sub-Microdegree Phase Measurement
Technique", submitted for publication in Review of Scientific Instruments,
American Institute of Physics.
\end{list}

\section{\large \bf Figure captions} Fig. 1: The function $\Phi(x)$
versus $x$, where $x=\frac{d}{r}$.
\\
\\ Fig. 2: The absolute phase shift $\phi(r)$ for $r$ between $d$ and
$1[m]$, at fixed parameter values  $d=0.01[m]$,
$\omega=251.3[rad/sec]$, and  $c=3\cdot10^{8}[m/sec]$.
\\
\\ Fig. 3: The change in phase shift $\Delta\phi$ for $r_{2}$ between
$r_{1}$ and $0.1[m]$, and for fixed parameter values $r_{1}=0.03[m]$,
$d=0.01[m]$,
$\omega=251.3[rad/sec]$, $c=3\cdot10^{8}[m/sec]$. Note that the value
$r_{1}=0.03[m]$ corresponds to the smallest possible separation between
the centers of the source mass and the detector mass in our experiment,
due to the finite size of the physical systems being used.

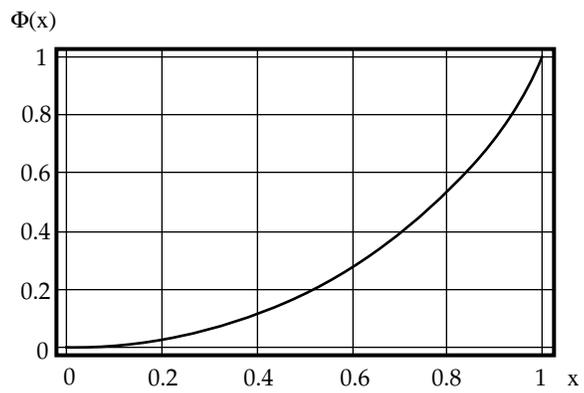
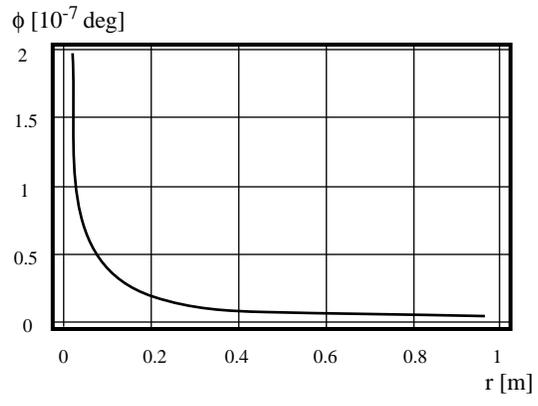
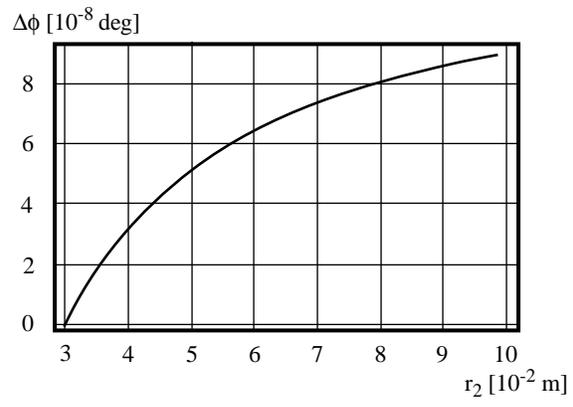
\end{document}